\documentclass[aps,prd,amsmath,amssymb,superscriptaddress,altaffillsymbol, preprintnumbers,nofootinbib,twocolumn]{revtex4}
\pdfoutput=1
\usepackage{amsthm}
\usepackage{graphicx}
\usepackage{float}
\usepackage{subfigure}
\usepackage{tikz}
\usepackage{color}
\newcommand{\beq}{\begin{eqnarray}}
\newcommand{\eeq}{\end{eqnarray}}

\newcommand{\bmp}{\noindent\begin{minipage}{16cm}}
\newcommand{\emp}{\end{minipage}\vskip 7mm} 

\usepackage{dcolumn}
\usepackage{bm}
\usepackage{bbm}
\usepackage{subfigure}
\usepackage{slashed} 

\usepackage{lipsum}

\theoremstyle{definition}

\theoremstyle{plain}

\usepackage{epsfig}

\usepackage{hyperref}
\definecolor{rossoCP3}{cmyk}{0,.88,.77,.40}
\definecolor{verdeCP3}{rgb}{0.09765625, 0.57421875, 0.1015625}
\definecolor{bluCP3}{rgb}{0, 0.23, 0.67}
\hypersetup{colorlinks, bookmarksopen, bookmarksnumbered,citecolor=bluCP3, linkcolor=bluCP3, pdfstartview=FitH, urlcolor=bluCP3}
\def\lsim{\mathrel{\rlap{\lower4pt\hbox{\hskip1pt$\sim$}}
    \raise1pt\hbox{$<$}}}                
\def\gsim{\mathrel{\rlap{\lower4pt\hbox{\hskip1pt$\sim$}}
    \raise1pt\hbox{$>$}}}                

\newcommand{\bea}{\begin{eqnarray}}
\newcommand{\eea}{\end{eqnarray}}

\newcommand{\ba}{\begin{eqnarray}}
\newcommand{\ea}{\end{eqnarray}}

\newcommand{\be}{\begin{eqnarray}}
\newcommand{\ee}{\end{eqnarray}}

\begin{document}
\title{Daily modulation and
gravitational focusing in direct dark matter search experiments}
%
\author{Chris Kouvaris}
\email{kouvaris@cp3-origins.net} 
\author{Niklas G. Nielsen}
\email{ngnielsen@cp3-origins.net} 
\affiliation{{\color{black} CP$^{3}$-Origins} \& Danish Institute for Advanced Study {\color{black} DIAS}, University of Southern Denmark, Campusvej 55, DK-5230 Odense M, Denmark}


\begin{abstract}
We study the effect of gravitational focusing of the earth on dark matter. We find that the effect can produce a detectable diurnal modulation in the dark matter signal for part of the parameter space which for high dark matter masses is larger than the diurnal modulation induced by the fluctuations in the flux of dark matter particles due to the rotation of the earth around its own axis. The two sources of diurnal modulation have  different phases and can be distinguished from each other. We demonstrate that the diurnal modulation  can potentially check the self-consistency of experiments that observe annual modulated signals that can be attributed to dark matter. Failing to discover a daily varying signal can result conclusively to the falsification of the hypothesis that the annual modulation is due to dark matter. We also suggest that null result experiments should check for a daily modulation of their rejected background signal with specific phases. A potential discovery could mean that dark matter collisions have been vetoed out. 
\preprint{CP3-Origins-2015-015 DNRF90, DIAS-2015-15.}
 \end{abstract}

\maketitle

\section{Introduction}
Direct detection of dark matter (DM) particles requires sophisticated ways of distinguishing a potential DM signal from noise that is due to other not so exotic particles such as neutrons, muons, photons etc. Most of direct detection experiments rely on vetoing signals that are consistent with the background. Therefore the correct and precise rejection of the background signal can make the difference between discovery of DM or not. Needless to say that although experiments have made enormous progress in understanding the respective backgrounds, one cannot be absolutely sure that the appropriate signal has been vetoed. However, there are other ways to circumvent the background problem. Experiments like DAMA rely on a different way of probing DM. Instead of rejecting background signals, experiments like these are looking for an annually modulated signal which could be consistent with DM. The earth is moving with respect to the reference frame  of the DM halo of our galaxy with a velocity that is mainly the velocity of the solar system superimposed by a much smaller component, i.e. the velocity of the earth around the sun. A variation in the velocity of the earth with respect to the DM halo leads to a variation of the flux of DM particles that reach an earth based detector, thus causing the aforementioned annual modulation of the signal. If no background component has a similar annual modulation, the annual modulation is solely attributed to DM. In principle it does not matter if someone is agnostic regarding the experimental background, since the existence of an annually modulated signal with a phase expected from the simplest models of the dark halo would be a strong hint that this is only a DM effect. Based on this principle, DAMA has observed an annual modulated signal that is consistent with the existence of DM~\cite{Bernabei:2013xsa}.

However there can be more sources of modulation in the DM signal. An obvious one is the angular velocity of the earth around its own axis. The earth is rotating around its own axis and therefore, the net velocity of the earth with respect to the DM halo fluctuates daily~\cite{Lee:2013xxa}. Given that the earth is moving with an average speed of $v_0=220 \text{km}/\text{s}$ (in the local standard of rest), the fluctuation due to the angular velocity of the earth is tiny, since the speed of a point on the equator is $\sim .5  \text{km}/\text{s}$. If the detected signal from DAMA is due to DM-detector collisions, one should at some point be able to observe also the diurnal modulation of the DM signal. This is an important self-consistency check because if DAMA observes an annual modulation and does not observe at some point a diurnal one, this is inconsistent and an alternative explanation of the annual modulated signal must be found. This is true for all experiments that base their detection technique on annual modulated signals. In fact recently, the DAMA collaboration studied the aforementioned effect concluding that it has not reached yet the required exposure to observe the diurnal modulation of the DM signal due to the angular velocity of the earth~\cite{Bernabei:2014jnz}. Another source of diurnal modulation  relevant for low DM masses is due to  stopping via DM interactions with underground nuclei~\cite{Kouvaris:2014lpa}. Depending on the distance that DM particles travel underground in order to reach the detector, they lose different amounts of energy due to interactions with  atoms and nuclei. Due to the rotation of the earth, this  creates a small amount of diurnal modulation signal. This can be particularly important in the case where DM is in the form of Strongly Interacting Massive Particles~\cite{Collar:1992qc,Hasenbalg:1997hs} or mirror dark matter~\cite{Foot:2011fh,Foot:2014osa}. Moreover, diurnal modulated DM signals can emerge from channelling effects in the detectors, since DM-nuclei scatterings from specific directions might have no quenching~\cite{Bozorgnia:2010zc,Bozorgnia:2011tk}.
Additionally, there can be a modulation in the DM signal due to gravitational focusing~\cite{Ling:2004aj}. Massive objects can work as a gravitational lens, changing the DM phase space density around them. This effect has been studied in the context of the sun acting as a lens, thus producing an annual modulated signal since the position of the earth with respect to the DM flux and the sun changes during the year~\cite{Bozorgnia:2014dqa, Lee:2013wza}. One should note here that although this amount of annual modulation is much smaller compared to the one due to the velocity of the earth revolving around the sun, it has a different phase and therefore it could potentially be distinguishable. Similarly, monthly modulated signals are also probable due to the focusing effect of the moon as well as the monthly varying velocity of the earth around the center of mass of the earth-moon system~\cite{Britto:2014wga}.

\begin{figure}[h]
\includegraphics[width=.5\textwidth]{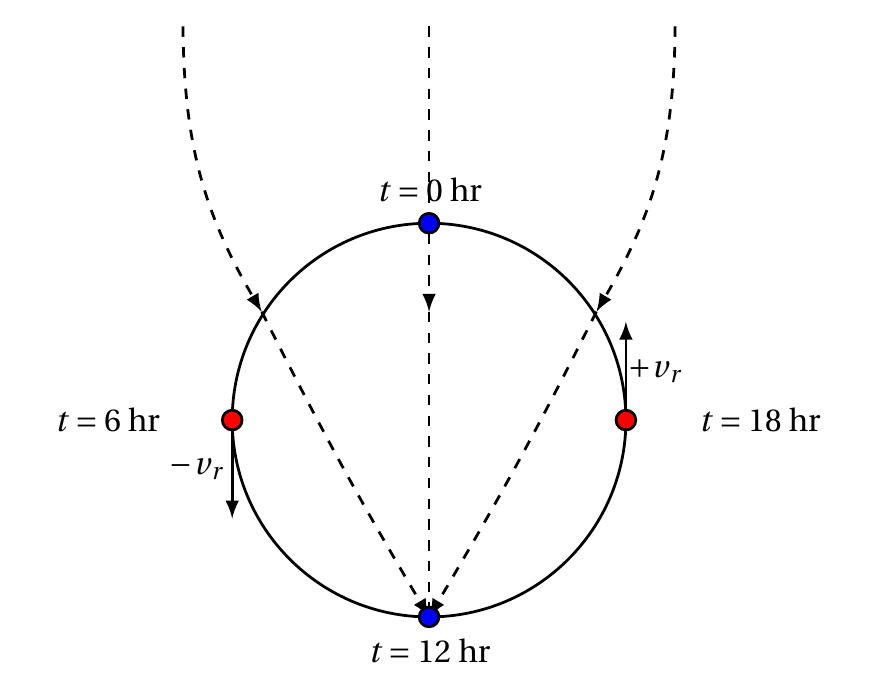}
\caption{This cartoon illustrates the two diurnal modulation signals of a detector rotating in the dark matter wind (dashed lines). We choose $t=0$ hr to be the time where the detector is maximally aligned against the dark matter wind. The red points are the extrema of the angular velocity modulation. The blue points at $t=0$ hr and $t=12$ hr are the minimum and maximum of the focusing modulation.}
\label{fig:cartoon}
\end{figure}

In this paper we study the effect of gravitational focusing produced by the earth itself acting as a lens. This produces a diurnal modulated signal because as the earth rotates around itself, there is a different amount of focusing depending on the relative position of the detector with respect to the center of the earth and the direction of the DM wind (i.e. the direction where the velocity of the earth with respect to the  reference system of the DM halo is maximum, see Fig.~\ref{fig:cartoon}). We find that the effect can be significant when the DM mass is large. Assuming spin-independent contact interactions and elastic scattering the DAMA annual modulated signal has roughly two regions in the cross section-DM mass parameter space that are consistent with DM, i.e. a low mass region around 10 GeV and a heavier one $\sim 50$ GeV. We find that the diurnal modulation due to earth gravitational focusing is larger than that produced by the angular velocity of the earth for part of the DM mass parameter space and therefore it should not be ignored. As we will demonstrate the diurnal modulated signal due to focusing is currently ``scratching" the allowed parameter space of DAMA in the large DM mass region. We should also emphasize that the two sources of diurnal modulation produce signals with different phases, therefore they can easily be identified once sufficient exposure has been reached.

In the next section we review the effect of gravitational focusing providing the relevant formulas. In section III, we calculate the diurnal modulation for gravitational focusing and due to the angular velocity of the earth, providing the amount of expected counts. We conclude in section IV.

\section{Gravitational Focusing}
The distortion of a distribution of particles due to the presence of a massive object has been studied  analytically~\cite{Danby1,Danby2,Griest:1987vc,Sikivie:2002bj,Alenazi:2006wu}. Here we are going to follow the derivation of~\cite{Sikivie:2002bj} the validity of which has been verified by numerical simulations~\cite{Alenazi:2006wu}.

Let us assume that DM follows a velocity distribution $f(\vec{v}_{\infty})$ in the absence of any focusing effect. The distribution is obviously independent of the position. Liouville's theorem states that the phase space density is conserved along particle trajectories. This means that the distribution of DM in the presence of a massive object (e.g the earth) should be

\begin{equation}
f_{e}(\vec{r},\vec{v})=f(\vec{v}_\infty(\vec{r},\vec{v})),
\label{eq:5.2}
\end{equation} 
where  $\vec{v}_\infty(\vec{r},\vec{v})$ is the velocity far from the gravitational source of the trajectory that passes by the position $\vec{r}$ with velocity $\vec{v}$. We briefly review  how one obtains $\vec{v}_\infty$ following~\cite{Sikivie:2002bj}. In order to obtain this velocity one should use the known constants of motion along the trajectory of the particle from a far away position where the gravitational field of the earth is negligible to the point where it enters the earth. The constants of motion are the energy, the angular momentum and the Laplace-Runge-Lenz vector per unit DM particle mass:
\bea
E & = & \frac{1}{2}v^{2} - \frac{G M_{e}}{r}\nonumber \\
\vec{l} & = & \vec{r} \times \vec{v} \nonumber \\ 
\vec{A} & = & \vec{v} \times (\vec{r} \times \vec{v}) 
- G M_{e} \hat{r}~~\ ,
\label{eq:5.3}
\eea
where $M_e$ is the mass of the earth. Setting the values of $E$, $\vec{l}$ and $\vec{A}$ at asymptotically far away distance equal to the corresponding values at distance $\vec{r}$ from the center of the earth, one finds~~\cite{Sikivie:2002bj}
\begin{widetext}
\beq
\vec{v}_\infty(\vec{r},\vec{v})=\frac{1}{a^{2}v_\infty^{2}+l^{2}}
\left\{ \vec{v} \left[ l^{2}-av_\infty^{2}r-av_\infty(\vec{r}\cdot\vec{v})
\right] + \vec{r}av_\infty^{2} \left[ \frac{1}{r}(\vec{r}\cdot\vec{v})
+\frac{v^{2}}{v_\infty}-\frac{a v_\infty}{r} \right] \right\}
\label{eq:5.8}
\eeq
\end{widetext}
where
$l^{2}=r^{2}v^{2}-(\vec{r}\cdot\vec{v})^2$, $a=GM_e/v_{\infty}^2$, $v_\infty$ is given by
\beq
v_\infty = \sqrt{v^{2}-\frac{2 G M_{e}}{r}}.
\label{eq:5.5}
\eeq
  We use a truncated Maxwell-Boltzmann DM distribution with $v_\infty< v_\text{esc}$ where $v_\text{esc}$ is the escape velocity from the galaxy. At asymptotically far away distance from the earth it takes the form
\beq
f(\vec{v}_\infty) = \frac{1}{N_\text{esc}(\pi v_0^2)^{3/2}}
\exp[-\frac{1}{v_0^{2}}(\vec{v}_\infty+\vec{v}_{e})^{2}].
\label{eq:5.1}
\eeq
Using Eq. (\ref{eq:5.2}) we get 
\beq
f_{e}(\vec{r},\vec{v})=\frac{1}{N_\text{esc}(\pi v_0^2)^{3/2}}
\exp[-\frac{v_{G}^{2}(\vec{r},\vec{v})}{v_0^2}],
\label{eq:5.9}
\eeq
where $N_\text{esc} = \text{erf}\left(v_\text{esc}/v_0\right) - 2v_\text{esc}\exp(-v_\text{esc}^2/v_0^2)/(\sqrt{\pi}v_0)$ and $\vec{v}_e$ is the velocity of the earth with respect to the rest frame of the DM halo. The velocity $v_{G}$ is given by
\begin{widetext}
\bea
v_{G}^{2}(\vec{r}, \vec{v}) & = &
(\vec{v}_{e}+\vec{v}_\infty(\vec{r},\vec{v}))^{2} 
=  \vec{v}_{e}^{2}+\vec{v}_{\infty}^{2} + 
2v_{e}\frac{1}{a^{2}v_\infty^{2}+l^{2}} \cdot\nonumber\\
& \cdot & \left\{v\cos\delta\left[l^{2}-av_\infty^{2}r-a 
v_\infty(\vec{r}\cdot\vec{v})\right]
+a v_\infty^{2}\cos\beta\left[\vec{r}\cdot\vec{v}
+\frac{v^{2}r}{v_\infty}-a v_\infty\right]\right\}~~~,
\label{eq:5.10}
\eea
\end{widetext}
where $\delta$ is the angle between $\vec{v}_{e}$  and the DM
velocity $\vec{v}$ and $\beta$ is the angle between $\vec{v}_{e}$ and
the position $\vec{r}$ of the observer relative to the center of the earth.
We will take $\vec{r}$ to be a point on the earth's surface. After a particle has reached the surface of the earth, it will move underground until it reaches the detector. However for this part of the trip, the above equations do not hold. This is because  the derivation of distribution (\ref{eq:5.9}) describes gravitational focusing around a point mass $M_e$. After the particle has entered the earth, one cannot treat  the earth as a point mass. Additionally, the Laplace-Runge-Lenz vector of Eq.~(\ref{eq:5.3}) is no longer conserved because inside the earth the gravitational force does not scale anymore as $\sim r^{-2}$ which is a necessary condition for  Eq.~(\ref{eq:5.3}) to be conserved. For the remaining part of the DM particle trajectory inside the earth we assume a constant mass density which corresponds to a gravitational force scaling as $\sim r$. For such a force there is a new Laplace-Runge-Lenz vector that is conserved. In the appendix, by applying conservation of energy, angular momentum and the new Laplace-Runge-Lenz vector, we demonstrate that in the undeground part of the particle's trip, no significant enhancement in the focusing takes place. Therefore, it is an excellent approximation to track the trajectory of the particle from asymptotically far distances to the surface of the earth and then assume that the particle travels in a straight line until it reaches the detector. In any case our approximation lies on the conservative side i.e. the actual focusing can only be larger than that calculated within our approximation since we ignore the focusing effect for the part of the particle's trajectory inside the earth. As we argue in the appendix, this makes a negligible difference in practice.

In order to compute how gravitational focusing of the earth can induce a diurnal modulation of the DM signal we follow the convention of~~\cite{Kouvaris:2014lpa} and introduce  
\be
\hat{n}=\hat{x} \cos \theta_l \cos\omega t+ \hat{y}\cos\theta_L \sin\omega t \pm \hat{z} \sin\theta_l, \label{nhat}
\ee
\be
\hat{v}_e=\hat{x} \sin \alpha+ \hat{z} \cos \alpha, \label{velhat}
\ee
where $\hat{n}$ is the unit vector with direction from the center of the earth to the detector, $\vec{v}_e$ is the velocity of the earth in the rest frame of the galaxy, $\theta_l$ is the latitude of the detector, and $\alpha$ the angle between $\vec{v}_e$ and  the angular velocity $\vec{\omega}$ of the earth. We have chosen a galactic rest frame where the $z$-axis is along the north-south pole, and $\vec{v}_e$ lies along the $x-z$ plane. It is understood that the $\pm$ corresponds to the north and south hemisphere. We have chosen $t=0$ the time where $\vec{v}_e$ and $\hat{n}$ align as much as possible i.e. $\hat{n}$ is along the $x-z$ plane. Recall that for the evaluation of Eq.~(\ref{eq:5.10}) we need to know both angles $\delta$ and $\beta$. By writing the velocity of the DM particle in spherical coordinates
\be
\vec{v}=v(\hat{x} \sin\theta \cos\phi +\hat{y} \sin\theta \sin\phi + \hat{z}\cos\theta), \label{vhat}
\ee
we can express the angle $\delta$  as 
\be
\cos\delta=\hat{v} \cdot \hat{v}_e=\sin\alpha \sin\theta \cos\phi +\cos\alpha \cos\theta. \label{delta}
\ee
In order to estimate the angle $\beta$ (defined as $\cos \beta= \hat{r} \cdot \hat{v}_e$), we need to associate the point where the particle crosses the surface of the earth with coordinates $\vec{r}$ with other known angles. From the geometry of the problem and upon assuming that particles move on straight lines after they have crossed the surface of the earth, it is easy to see that 
\be
\hat{r}=\frac{\ell}{R_{\oplus}} \hat{v}+\frac{R_{\oplus}-\ell_D}{R_{\oplus}} \hat{n}, \label{hatr}
\ee
where $R_{\oplus}$ is the radius of the earth, $\ell_D$ the depth of the detector, and $\ell$ the length of the trajectory (i.e. the length of the straight line) from the point where the particle enters the earth to the detector. This length is given by 
\bea
 \ell &=& \left( R_{\oplus}-\ell_{D} \right)\cos \psi  \label{length} \\ \nonumber 
&+& \sqrt{(R_{\oplus}-\ell_{D})^{2} \cos^{2} \psi - (\ell_{D}^{2} - 2 R_{\oplus}\ell_{D})}, 
\eea
where  $\psi$ is the angle between the velocity of the particle and $\hat{n}$. From Eqs.~(\ref{nhat}) and (\ref{vhat}) we get
\begin{eqnarray}
\cos\psi=\hat{v}\cdot \hat{n}=\cos\theta_l \cos\omega t \sin \theta \cos\phi \nonumber\\
+ \cos\theta_l \sin\omega t \sin\theta \sin\phi \pm \sin\theta_l \cos\theta. \label{psii}
\end{eqnarray}
Using Eq.~(\ref{hatr}) we get
\be
\cos \beta= \hat{r} \cdot \hat{v}_e=\frac{\ell}{R_{\oplus}} \cos \delta +\frac{R_{\oplus}-\ell_D}{R_{\oplus}} \hat{n}\cdot \hat{v}_e.
\ee
From Eqs.~(\ref{nhat}) and (\ref{velhat}) is trivial to deduce $\hat{n}\cdot \hat{v}_e$ and the final expression for $\cos \beta$ becomes
\begin{align}
\cos \beta&=\notag\\
 &\frac{\ell}{R_{\oplus}} \cos \delta +\frac{R_{\oplus}-\ell_D}{R_{\oplus}} (\sin \alpha \cos\theta \cos \omega t \pm \cos \alpha \sin \theta_l).
\end{align}
One can see from Eq.~(\ref{eq:5.10}) that $v_G$ depends on $\beta$ which depends on time. This is what creates the diurnal modulation in the DM signal.

\begin{figure*}
\includegraphics[width=.8\textwidth]{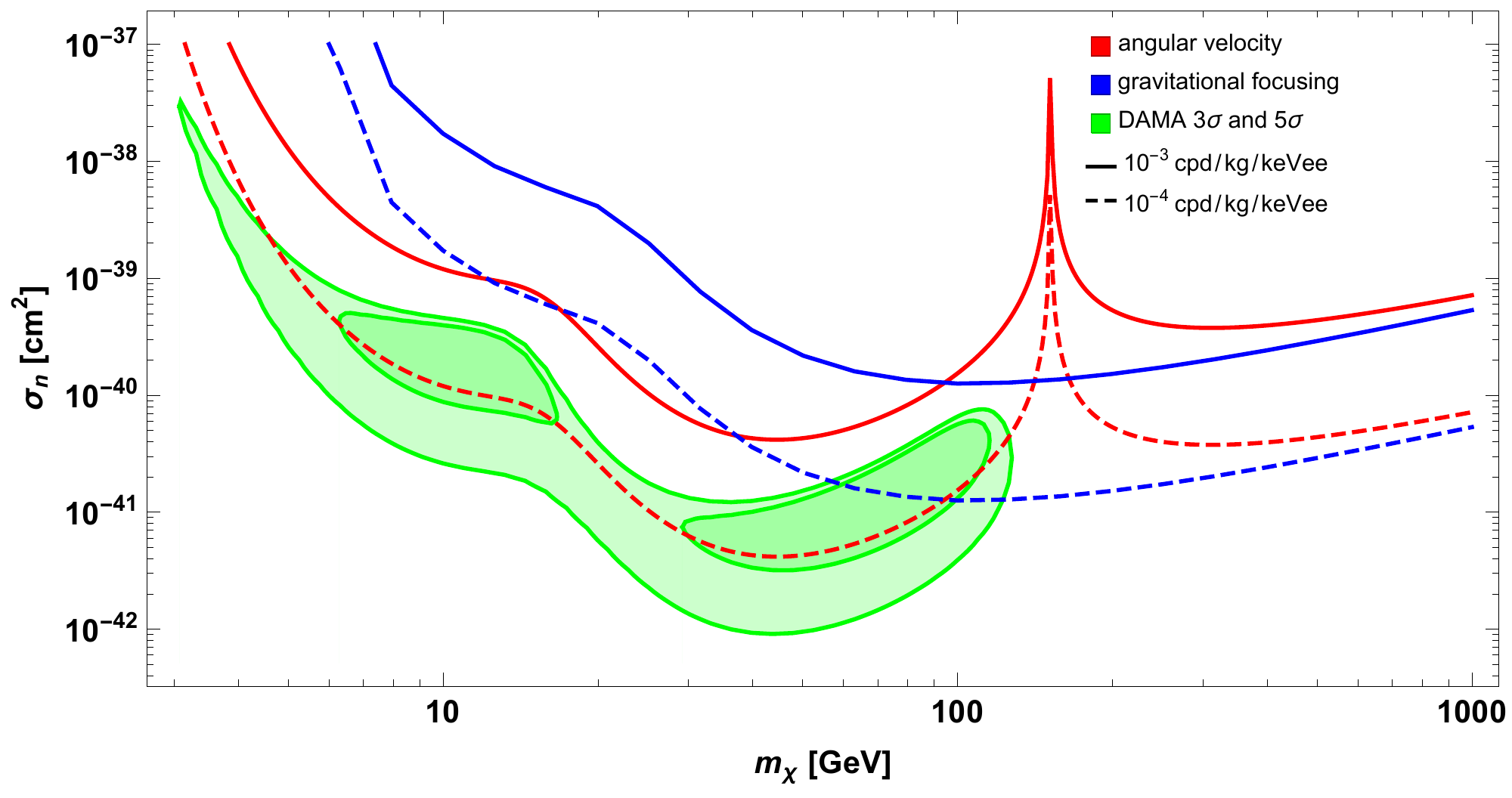}
\caption{The solid (dashed) lines correspond to a modulation amplitude of $10^{-3}$ ($10^{-4}$) cpd/kg/keVee in the 2-6 keV bin for DAMA. The red lines correspond to the modulation induced by the angular velocity of the earth, whereas the blue ones correspond to gravitational focusing.  The green contours are the region favoured by DAMA at $3\sigma$ and $5\sigma$.}
\label{fig:mainfig}
\end{figure*}

\section{Diurnal Modulation}
The differential rate of dark matter events in a detector is given by
\begin{equation}
\frac{dR}{dE_R} = \sum_i N_{T,i}\frac{\rho_\chi}{m_\chi}\int_{v_{\mathrm{min}}}^{v_\text{esc}+v_e}v \frac{d\sigma}{dE_R} f_e(\vec{r},\vec{v})d^3\vec{v}
\label{eq:rate}
\end{equation}
where $m_\chi$ is the DM mass, $\rho_\chi = 0.3$ GeV/cm$^3$ is the DM density, $v_\text{esc} = 550$ km/s is the galactic escape velocity and $N_{T,i}$ is the number of nuclei targets of type $i$ in the detector. For example in the case of DAMA, the detector is made of   NaI  where $N_{T,\mathrm{Na}} = N_{T,\mathrm{I}}= M_{\mathrm{tot}}/(m_{\mathrm{Na}}+m_{\mathrm{I}})$ ($M_{\mathrm{tot}}$ being the total mass of the detector). $f_e(\vec{r},\vec{v})$ is given in Eq. (\ref{eq:5.9}), $v_{\mathrm{min}}=\sqrt{m_N E_R/2\mu^2}$, and $d\sigma/dE_R$ is the differential DM-nucleus cross section, $m_N$ and $\mu$ being the nucleus mass and the reduced mass of DM and nucleus respectively.

The diurnal modulation caused by the angular velocity of the earth spinning around its own axis  enters through the time dependence of  $\vec{v}_e$ in Eq.~(\ref{eq:5.1}). $\vec{v}_e$ is 
\begin{equation}
\vec{v}_e(t) = \vec{v}_\odot + \vec{v}_{\mathrm{rev}}(t)+\vec{v}_{\mathrm{rot}}(t),
\end{equation}
where $\vec{v}_\odot$ is the velocity of the sun in the galactic frame, $\vec{v}_{\mathrm{rev}}(t)$ is earth's velocity as it revolves annually around the sun, and $\vec{v}_{\mathrm{rot}}(t)$ is the rotational velocity of the earth around its own axis. Since $\vec{v}_{\mathrm{rev}}(t)$ varies slowly and is small compared to $\vec{v}_\odot$ the term relevant on daily time scales is $\vec{v}_{\mathrm{rot}}(t)$. We will take $\vec{v}_{\mathrm{rev}}(t) =0$ since it will only contribute to the constant part of the velocity during any particular daily cycle. In absolute values we write:
\begin{equation}
v_e(t) \simeq v_\odot +  V_{\mathrm{rot}} \cos \left[\omega (t-t_{\mathrm{rot}})\right],
\end{equation} 
where $v_\odot =232 \pm 50$ km/s,  $\omega = 2\pi/24$ hr$^{-1}$ and $t_{\mathrm{rot}} = 18$ hr. Note that our $t=0$ corresponds to the time where the detector aligns maximally against the DM wind. $V_{\mathrm{rot}} =\omega R_{\oplus} \cos\theta_l \cos\kappa =0.23$ km/s which is the speed due to the rotation of the earth at latitude $\theta_l=42$ degrees,  $\kappa$ being the angle between $\vec{v}_{\text{rot}}$ and  $\vec{v}_\odot$~\cite{Bernabei:2014jnz}.

We assume that scattering on detector nuclei is described by a contact interaction 
\begin{equation}
\frac{d\sigma}{dE_R} = \frac{m_N \sigma_p A^2}{2\mu_p^2v^2}F^2(E_R),
\label{eq:cross section}
\end{equation}
where we provide the cross section in terms of the normalized DM-nucleon cross section $\sigma_p$.  $\mu_p$ is the DM-nucleon reduced mass, $A$ is the number of nucleons in the nucleus and $F(E_R)$ is nuclear form factor. If one neglects gravitational focusing for the moment, the velocity integral of Eq. (\ref{eq:rate}) can be calculated analytically
\begin{equation}
\eta(E_R,t) = \int_{v_{\mathrm{min}}}^{v_{\mathrm{esc}}+v_e} \frac{f(\vec{r},\vec{v})}{v}d^3\vec{v}.
\label{eq:etaInt}
\end{equation}
The result of this integral is~\cite{Savage:2006qr}:
\begin{widetext}
\begin{equation} 
  \eta(E_R,t) =
    \begin{cases}
      \frac{1}{2 N_{\mathrm{esc}} v_0 y}
        \left[
          \mathrm{erf}(x\!+\!y) - \mathrm{erf}(x\!-\!y) - \frac{4}{\sqrt{\pi}} y e^{-z^2}
        \right] \, ,
        & \textrm{for} \,\, z>y, \, x<|y\!-\!z| \\
      \frac{1}{2 N_{\mathrm{esc}} v_0 y}
        \left[
          \mathrm{erf}(z) - \mathrm{erf}(x\!-\!y) - \frac{2}{\sqrt{\pi}} (y\!+\!z\!-\!x) e^{-z^2}
        \right] \, ,
        & \textrm{for} \,\, |y\!-\!z|<x<y\!+\!z \\
      0 \, ,
        & \textrm{for} \,\, y\!+\!z<x
    \end{cases}
    \label{eq:eta}
\end{equation}
\end{widetext}
where $x= v_{\mathrm{min}}/v_0$, $y= v_e(t)/v_0$ and $z=(v_{\mathrm{esc}}+v_e)/v_0$.

Although our formalism is generic and can be valid for any experiment, we are going to focus here on DAMA since it is the experiment with a confirmed annual modulation signal with more than $9\sigma$ confidence when considering the full exposure of 1.33 ton $\times$ yr. The DAMA collaboration has looked for diurnal modulated signals in~\cite{Bernabei:2014jnz} where an upper bound for the amplitude of diurnal modulation has been imposed, albeit only considering the effect of the angular velocity of the earth and not the gravitational focusing. In this paper DAMA has presented the experimental diurnal residual rate of the single-hit scintillation events, in different energy bins
 as a function of the hour of either a solar or a sidereal day.  The cumulative exposure analysed in~\cite{Bernabei:2014jnz} is 1.04 ton $\times$ yr. In order to compare with the registered counts of DAMA we use
\be
\frac{dR}{dE_R'}=\int_0^\infty \frac{dR}{dE_R} G(E_R',E_R)dE_R,
\label{eq:convolution}
\ee
where $G(E_R',E_R)=\exp[-(E_R'-qE_R)^2/2\sigma_e^2]/\sqrt{2 \pi}\sigma_e$ is a convolution function that takes into account the energy resolution of the experiment, $q$ is the quenching factor of the scattering nucleus (i.e. 0.3 for Na and 0.09 for I) and $\sigma_e$ characterizes the energy resolution of the experiment~\cite{Bernabei:2008yh}. 
The total rate in the $k$-th energy bin is then
\begin{equation}
R_k = \int_{E_k}^{E_{k+1}}  \frac{dR}{dE_R'} dE_R',
\end{equation}
and the modulation amplitude in cpd/kg/keVee is
\begin{equation}
A_k =\frac{1}{2} \frac{\max_{t} R_k - \min_{t} R_k}{E_{k+1}-E_k}.
\end{equation}

 Fig.~\ref{fig:cartoon} illustrates the fact that the maximal event rates for gravitational focusing and the rotation of the earth occur at different times. For gravitational focusing maximum and minimum rates occur at $t=12$ hr and $t=0$ respectively. For the angular velocity modulation amplitude in the 2-6 keVee energy bin, if DM mass is below 152 GeV, the maximum and minimum occur at $t=18$ hr and $t=6$ hr respectively simply because when the rotational velocity of the earth aligns maximally with $\vec{v}_{\odot}$, the DM flux maximizes. However, for DM masses above 152 GeV the maximum occurs at $t=6$ hr and the minimum at  $t=18$ hr. The reason for this phase shift is that although the overall DM flux is maximum at $t=18$ hr, Eq.~(\ref{eq:rate}) is dominated by low velocities.

Fig.~\ref{fig:mainfig} shows the DM-nucleon cross section vs DM mass parameter space assuming the distribution of Eq.~(\ref{eq:5.9}) and cross section of Eq.~(\ref{eq:cross section}). The lines of Fig.~\ref{fig:mainfig} show what part of the DM-nucleon cross section vs DM mass leads to diurnal modulation rates (in the 2 to 6 keVee bin of DAMA) of  $10^{-3}$ cpd/kg/keVee (solid lines) and  of $10^{-4}$ cpd/kg/keVee (dashed lines) for either gravitational focusing (blue lines) or angular velocity (red lines). The pole in the red curves corresponds to the flip in phase for the angular rotation effect where the modulation amplitude vanishes and an infinite cross section is needed to provide a finite modulation rate. For a sense of scale DAMA/Libra phase-1 has already excluded  modulations larger than $1.2 \times 10^{-3}$ cpd/kg/keVee at 90\% C.L. \cite{Bernabei:2014jnz} (although this limit is only on the angular velocity effect with a constant phase of $t_{\mathrm{max}}=18$ hr). One can see that gravitational focusing becomes the dominant source of modulation at large dark matter masses. In our plot we have also included the DAMA favoured region based on the observation of the annual modulation.

\begin{figure}
\includegraphics[width=.4\textwidth]{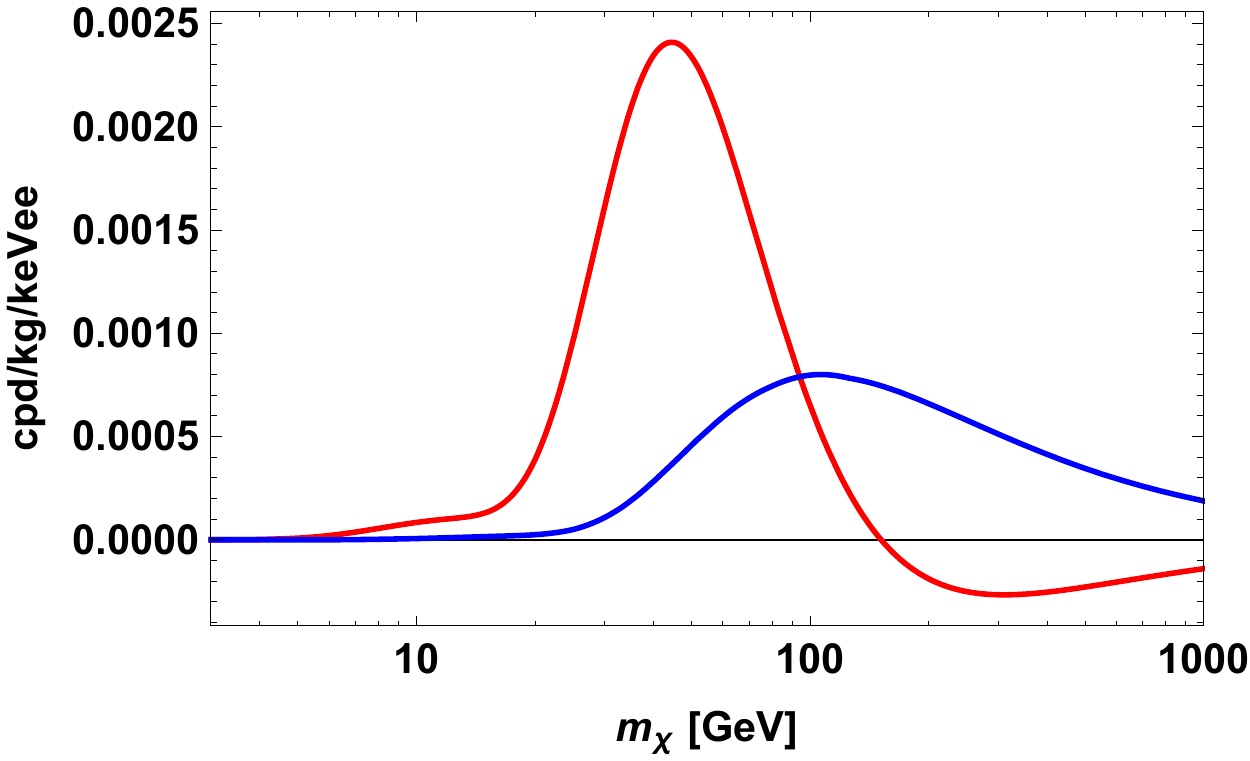}
\caption{The predicted modulation amplitude of daily modulation for the angular velocity effect (red) and gravitational focusing (blue), calculated at $\tilde{\sigma}_p = 10^{-40}$ cm$^2$. The amplitude for an arbitrary cross section is obtained by rescaling with a factor $\sigma_p/\tilde{\sigma}_p$.}
\label{fig:AmpCompare}
\end{figure}

\begin{figure}
\includegraphics[width=.4\textwidth]{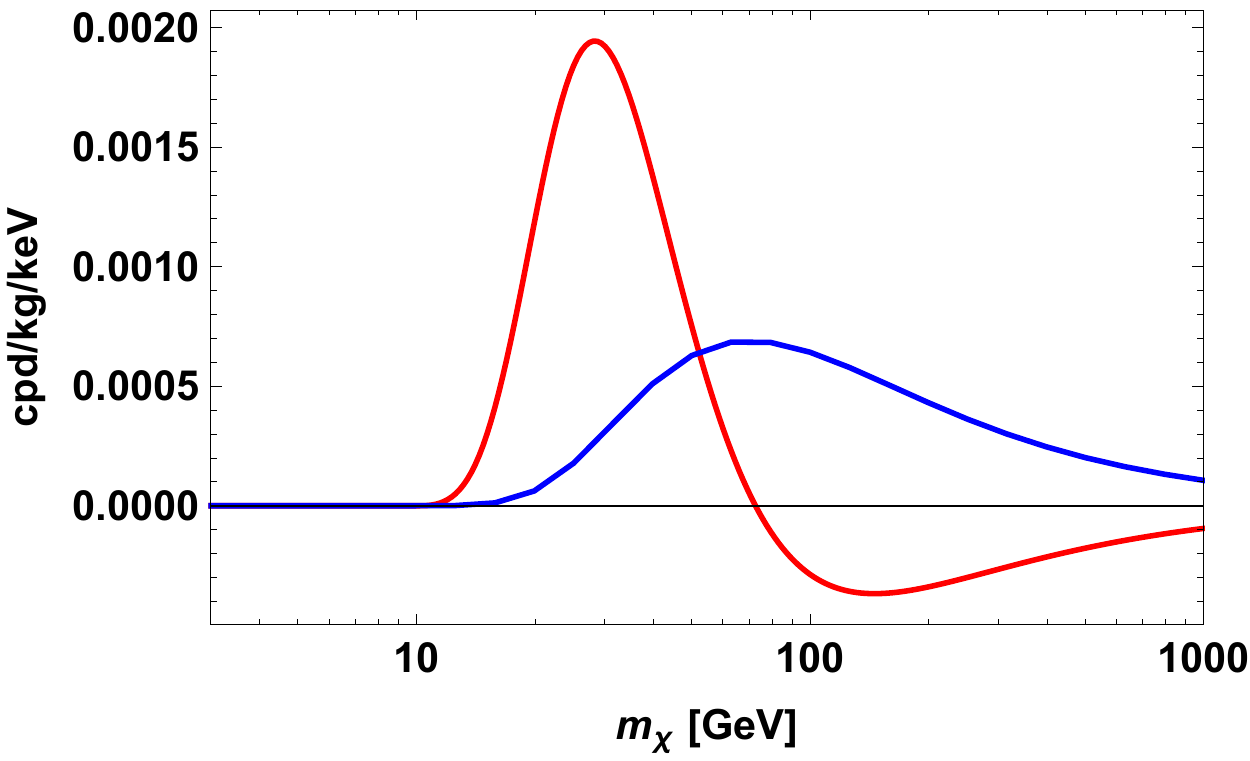}
\caption{The predicted modulation amplitude of daily modulation for a Xenon detector with perfect energy resolution, i.e. $G(E'_R,E_R)$ of Eq. (\ref{eq:convolution}) is a delta function $\delta(E_R -  E_R')$. We show the amplitude due to angular velocity  (red) and gravitational focusing (blue), calculated at $\tilde{\sigma}_p = 10^{-40}$ cm$^2$ in the 10-20 keV recoil energy bin. The amplitude for an arbitrary cross section is obtained by rescaling with a factor $\sigma_p/\tilde{\sigma}_p$.}
\label{fig:AmpCompareXe}
\end{figure}

In Fig.~\ref{fig:AmpCompare} we show the predicted modulation amplitude in cpd/kg/keVee in the 2-6 keVee bin of a NaI detector at a 42 degree latitude (the latitude of Gran Sasso), where a contact type interaction has been assumed.
 Although we present our results for a standard value of $\tilde{\sigma}_p=10^{-40}$ cm$^2$,  the modulation amplitude can be obtained for an arbitrary cross section simply by rescaling the number of counts in the plot with a factor $\sigma_p/\tilde{\sigma}_p$. At small DM masses ($\lesssim 100$ GeV), earth's  angular velocity is the dominant source of diurnal modulation. At large masses ($\gtrsim 100$ GeV) the gravitational focusing causes a higher diurnal modulation rate. We find that the overall interference is constructive in the region where gravitational focusing is sizeable. In Fig. \ref{fig:AmpCompareXe} we present the same results assuming a Xenon target.

\begin{figure}
\includegraphics[width=.4\textwidth]{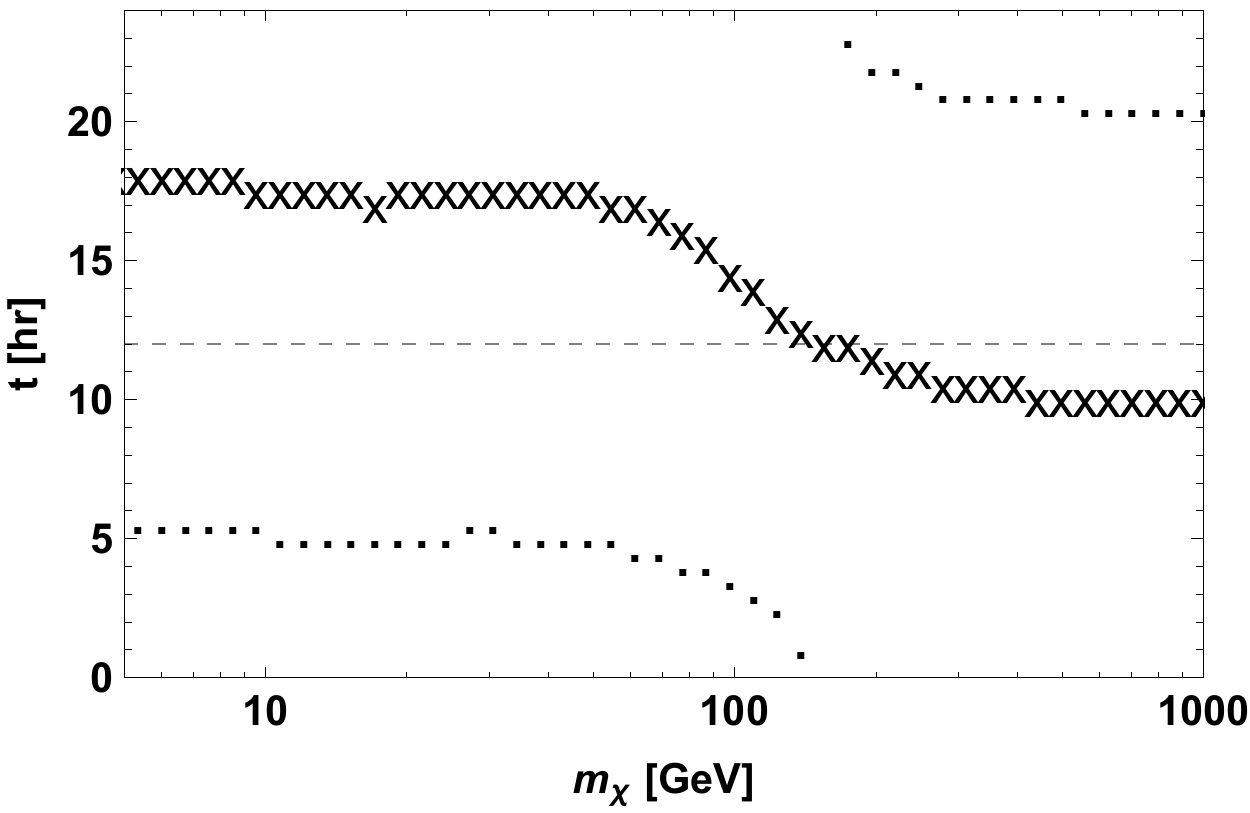}
\caption{Respective times where the maximum ($\times$) and the minimum  ($\mathbf{\cdot}$) number of counts occur as a function of the DM mass. The dashed line  $t=12$ hr  is the time
where  gravitational focusing maximizes (if considered alone).}
\label{fig:phases}
\end{figure}

As we have already mentioned the two sources of diurnal modulation we examined have generally different amplitudes and different phases. This is important because it is  possible to identify the two signals. In the cases where both signals are comparable, due to the difference in the phase, they can interfere either constructively or destructively. In this case the times where the maximum and minimum number of counts occur will not coincide with either one of the two individual ones (i.e. gravitational focusing and angular velocity). It will rather occur in different times according to how the two signals interfere. The times during the day where maximum and minimum number of counts are observed do not depend on the cross section but only on the DM mass and on the energy.
 In Fig.~\ref{fig:phases} we have
plotted the time at which the maximal and minimal rates are observed during the diurnal cycle. At small DM masses the angular velocity effect is dominant so the maximum rate takes place  at $t_{\mathrm{max}} = 18$ hr. At a DM mass of $\sim 100$ GeV gravitational focusing becomes the dominant effect and the maximum rate is found at $t_\mathrm{max} = 12$ hr. Finally when the phase of the angular velocity effect shifts at $m_{\chi}=152$ GeV, the combined $t_\mathrm{max}$ settles at around 10 hr. The minimal rate occurs approximately 12 hr after the maximal.

\begin{figure}
\includegraphics[width=.4\textwidth]{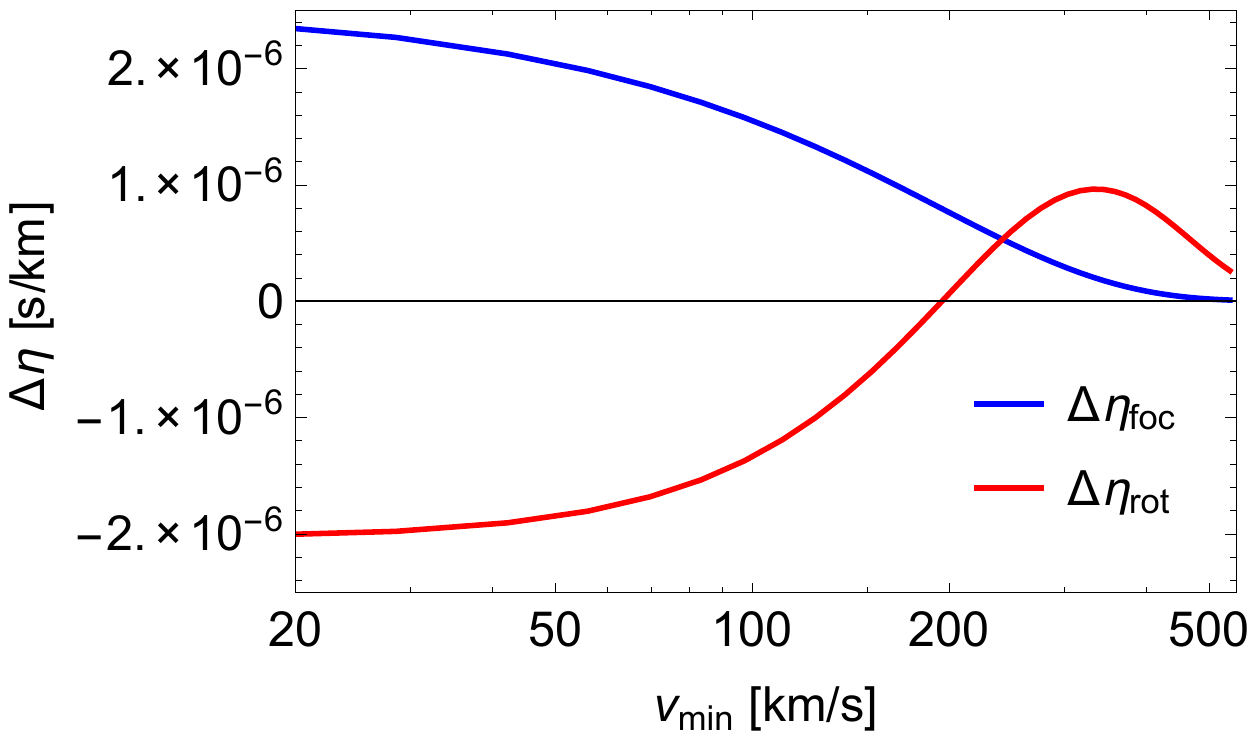}
\caption{The modulation of the velocity integral $\eta$ for the two effects. For low $v_\text{min}$ the focusing effect is larger. Around $v_\text{min} \approx 200$ km/s the rotation effect changes phase and becomes the most important effect when $v_\text{min} \gtrsim 240$ km/s.}
\label{fig:deltaeta}
\end{figure}

In addition to Fig.~\ref{fig:AmpCompare} and \ref{fig:AmpCompareXe} we show in Fig.~\ref{fig:deltaeta} the parameter 
\begin{equation}
\Delta\eta(v_\text{min}) =\frac{1}{2} \left[\max_{t}\eta(v_\text{min},t) - \min_{t}\eta(v_\text{min},t)\right],
\end{equation}
where $\eta$ is the velocity integral in Eq. (\ref{eq:etaInt}) and $\Delta \eta$ corresponds to the difference between the maximum and minimum value of $\eta$ in the course of a day. For the modulation induced by earth's rotation $\eta$ has the analytical expression of Eq. (\ref{eq:eta}), while for focusing we numerically integrate Eq. (\ref{eq:etaInt}) with the distribution of Eq. (\ref{eq:5.9}). The modulation of $\eta$ is more model independent than the rate, since it is not integrated over a specific energy range and only assumes a dependence $d\sigma/dE_R \propto v^{-2}$ and a Maxwell-Boltzmann velocity distribution. For the rotation effect we show $\Delta\eta_\text{rot}(v_\text{min}) = \left[\eta(v_\text{min},18\,\text{hr}) - \eta(v_\text{min},6\,\text{hr})\right]/2$ and for the gravitational focusing effect we show $\Delta\eta_\text{foc}(v_\text{min}) = \left[\eta(v_\text{min},12\,\text{hr}) - \eta(v_\text{min},0\,\text{hr})\right]/2$. The cusp point in Fig.~\ref{fig:mainfig} at $m_\chi = 152$ GeV corresponds to the $v_\text{min}$ where $\Delta\eta_\text{rot}(v_\text{min})$ crosses zero in Fig.~\ref{fig:deltaeta}. This can be checked by taking a 6 keV recoil on iodine which corresponds to $v_\text{min} \approx 208$ km/s. 

\section{Conclusions}
In this paper we studied two sources of diurnal modulation in the dark matter signal. The first is due to the gravitational focusing of the earth, i.e. the earth acts as a gravitational lens that changes the local dark matter phase space density. Due to the rotation of the earth around its axis, an earth based detector that is not placed at the poles of the earth will experience a daily variation in the amount of dark matter focusing. Additionally, the angular velocity of earth due to its revolution around its own axis causes a variation of the total velocity of the detector with respect to the rest frame of the dark matter halo, thus creating  a daily fluctuation of the dark matter signal. As we showed in this paper the phases of the two diurnal variations are distinct. For gravitational focusing the minimum signal takes place when the detector alignes as much as possible against the dark matter wind, while the maximum occurs 12 hours later. On the contrary the diurnal modulation due to the rotation velocity of the earth around its axis becomes maximum when the the rotation velocity aligns maximally against the dark matter wind. In the latter case for heavy enough dark matter masses, the phase shifts by $\pi$ as we explained in the text. We found that for low dark matter masses the ``angular velocity" source of modulation is larger but the picture is reversed above a certain mass. 

We have provided the expected number of counts as well as the exact times where the maximum and the minimum signal should be observed for any mass and cross section. This could be quite important for two reasons: Experiments like DAMA that observe annual modulated signals and therefore are compatible to dark matter signals should pay attention to the diurnal signal. Failing to find it would mean that there is no self-consistency and the annual modulation is not caused by dark matter. It is an important self-consistency check. On the other hand, we suggest  null result experiments like Xenon and LUX to check their rejected background for signs of diurnal modulation. Such a potential discovery would mean that  dark matter signals might resemble for example electron events and have been vetoed  until now.  \\ \\
{\bf \hspace{3cm} Acknowledgements}
We would like to warmly thank I. Shoemaker for valuable help and comments. C.K. would like to thank the Munich Institute for Astro- and Particle Physics for its hospitality. Both authors are supported by the Danish National Research Foundation, Grant No. DNRF90. \\ 

{\bf \hspace{3cm} APPENDIX}\\
We examine here how good is the approximation we have made in the study of gravitational focusing by assuming that particles follow straight lines after they have crossed the surface of the earth towards the underground detector. In other words, we ignore the focusing effect for the part of the trajectory inside the earth. In order to figure out how good the approximation is let us assume that the earth has uniform density. For a particle entering the earth the gravitational force from the usual $1/r^2$ dependence becomes $F=-(4 \pi/3)G \rho m r$ where $\rho$ and $m$ are the uniform density of the earth and the mass of the dark matter particle.
For this linear in distance force, the Laplace-Runge-Lenz vector becomes~\cite{harm}
\begin{equation}
\vec{A}_e=\frac{1}{\sqrt{mr^2\omega B-mr^2 E+L^2}}\{\vec{p} \times \vec{L}+(mr\omega B-mrE)\hat{r}\},
\end{equation}
where $B=(E^2-\omega^2 L^2)^{1/2}/\omega$, $\omega=\sqrt{k/m}$, $k=4\pi G\rho m/3$ and $E$ and $L$ are the kinetic energy and angular momentum of the particle. $\vec{A}_e$ can be rewritten in the form
\begin{equation}
\vec{A}_e=C_1\hat{v} \times (\hat{r} \times \hat{v})+C_2 \hat{r},
\end{equation}
where $C_1=m^2 v^2r(mr^2\omega B-mr^2 E+L^2)^{-1/2}$ and $C_2=(mr\omega B-mrE)(mr^2\omega B-mr^2 E+L^2)^{-1/2}$. After using well known vector identities, $\vec{A}_e$ finally becomes
\begin{equation}
\vec{A}_e=(C_1+C_2)\hat{r}-C_1(\hat{v}\cdot\hat{r}) \hat{v}.
\end{equation}
Let us consider a particle crossing the surface of the earth at point 1 following a trajectory that ends in the detector at point 2. The conservation of $\vec{A}_e$ at the two points gives
\begin{equation}
\Gamma_1 \hat{r}_1-\Delta_1(\hat{v}_1 \cdot \hat{r}_1) \hat{v}_1=\Gamma_2 \hat{n}-\Delta_2(\hat{v} \cdot \hat{n}) \hat{v},\label{ape_1}
\end{equation}
where $\Gamma_{1,2}$ is $C_1+C_2$ evaluated at points 1 and 2 respectively. Note that we have dropped the index 2 from the vectors and that by definition $\hat{n}=\hat{r}_2$. $\Delta_{1,2}$ is $C_1$ evaluated at the corresponding points. Conservation of angular momentum in points 1 and 2 give
\begin{equation}
v_1R_{\oplus}\hat{r}_1 \times \hat{v}_1=v \ell_c \hat{r} \times \hat{v},
\end{equation}
where $\ell_c=R_{\oplus}-\ell_D$ is the distance of the detector from the center of the earth. The above equation gives 
\begin{equation}
\sin\psi_1=\frac{v \ell_c}{v_1 R_{\oplus}}\sin\psi, \label{psi}
\end{equation}
where $\psi_1$  ($\psi$) is the angle between $\hat{r}_1$ and $\hat{v}_1$ ($\hat{n}$ and $\hat{v}$). Since we want to determine how large is the effect of focusing in the trajectory during the flight of the particle inside the earth, let us take a trajectory where the velocity of the particle does not align with the direction of the detector $\hat{n}$. In this case $\vec{v}_1$ and $\hat{n}$ define a plane. Since the gravitational force is central, the trajectory of the particle will remain on this plane. Therefore we can express the direction of $\vec{v}_1$ as
\begin{equation}
\hat{v}_1=\epsilon_1 \hat{n}+\epsilon_2 \hat{v},\label{epsilon}
\end{equation}
where $\epsilon_{1,2}$ are coeffcients to be determined. A first equation regarding $\epsilon_{1,2}$ is obtained by demanding $\hat{v}_1\cdot \hat{v}_1=1$. This leads to 
\begin{equation}
\epsilon_1^2+\epsilon_2^2+2 \epsilon_1\epsilon_2 \cos\psi =1.
\end{equation}
We can rewrite now Eq.~(\ref{ape_1}) by substituting $\hat{r}_1$ by that of Eq.(\ref{hatr}) (since this is entry point of particle into the earth) and $\hat{v}_1$ by Eq.~(\ref{epsilon}) keeping in mind that $\hat{v}_1 \cdot \hat{r}_1=\cos\psi_1$ and $\hat{v} \cdot \hat{n}=\cos\psi$
\begin{widetext}
\begin{equation}
\left (\Gamma_2+\epsilon_1 \Delta_1 \cos\psi_1-\Gamma_1\frac{\ell_c}{R_{\oplus}} \right )\hat{n}= \left (\Delta_2 \cos\psi+\Gamma_1\frac{\ell}{R_{\oplus}}-\epsilon_2\Delta_1\cos\psi_1 \right ) \hat{v}.
\end{equation}
\end{widetext}
If we multiply both sides for example with $\hat{v}$ we get
\begin{widetext}
\begin{equation}
\left (\Gamma_2+\epsilon_1 \Delta_1 \cos\psi_1-\Gamma_1\frac{\ell_c}{R_{\oplus}} \right )\cos\psi= \left (\Delta_2 \cos\psi+\Gamma_1\frac{\ell}{R_{\oplus}}-\epsilon_2\Delta_1\cos\psi_1 \right ). \label{epsilon2}
\end{equation}
\end{widetext}
 Eqs.~(\ref{epsilon}) and (\ref{epsilon2}) form a system that can be solved in terms of $\epsilon_1$ and $\epsilon_2$. One can use Eq.~(\ref{psi}) to determine $\cos\psi_1$ in terms of $\cos\psi$ given by Eq.~(\ref{psii}). For a variety of different angles of entry for the projectile particle, we have numerically solved  Eqs.~(\ref{epsilon}) and (\ref{epsilon2}) and found that in all cases $\epsilon_2\sim 1$ and $\epsilon_1 \sim 0$. This means that $\hat{v}_1\simeq \hat{v}$ and therefore the approximation we have made that the particle moves on a straight trajectory for the part of the trip which takes place inside the earth is valid.

\end{document}